\documentstyle[pra,aps,amssymb,graphicx]{revtex}
\begin{document}
\draft

\def\overlay#1#2{\setbox0=\hbox{#1}\setbox1=\hbox to \wd0{\hss #2\hss}#1%
\hskip
-2\wd0\copy1}
\twocolumn[
\hsize\textwidth\columnwidth\hsize\csname@twocolumnfalse\endcsname

\title{Optics of Nonuniformly Moving Media}
\author{U.\ Leonhardt$^1$ and P.\ Piwnicki$^{1,2}$}
\address{~$^1$Physics Department, Royal Institute of Technology (KTH),
Lindstedtsv\"agen 24, S-10044 Stockholm, Sweden}
\address{~$^2$Abteilung f\"ur Quantenphysik, Universit\"at Ulm,
D--89069 Ulm, Germany}
\maketitle
\begin{abstract}
A moving dielectric appears to light as an effective gravitational field.
At low flow velocities the dielectric acts on light in the same way
as a magnetic field acts on a charged matter wave.
We develop in detail the geometrical optics of moving dispersionless media.
We derive a Hamiltonian and a Lagrangian to describe ray propagation.
We elucidate how the gravitational and the magnetic model of 
light propagation are related to each other.
Finally, we study light propagation around a vortex flow.
The vortex shows an optical Aharonov--Bohm effect at large 
distances from the core, and, at shorter ranges, the vortex may
resemble an optical black hole.
\end{abstract}
\date{today}
\vskip2pc]
\narrowtext

\section{Introduction}

Consider a glass container filled with a transparent liquid, say water. 
Let a plane wave of coherent laser light travel through the water. 
Obviously, not much will happen. 
The light will remain a plane wave and will only gather an overall 
phase shift. 
Now imagine that the water is set into motion. 
For example, a magnetic mixer at the bottom of the container creates a
vortex.
Let us assume that no air bubbles contaminate the transparent liquid 
and that no heat gradient is generated. 
We know that water is to a large degree incompressible. 
Therefore the refraction index of the whirling liquid is spatially uniform. 
Will the light remain a plane wave? 
Maybe surprisingly, it will not. 
Instead, the light will develop an interference structure that is
sensitive to the velocity of the liquid. 
Furthermore, if we send in a narrow
laser beam, the vortex will bend the ray.
A moving medium drags light.
This effect can be employed to gather information about 
the flow of a transparent liquid.
One could think of reconstructing an unknown velocity profile 
from measured interference patterns,
as a form of optical tomography.

In this paper we develop a systematic theory that explains motional
effects of a nondispersive dielectric medium on light propagation.
We postulate that the wave equation is valid in all locally comoving 
frames of the medium.
Then we transform the wave equation to the laboratory frame.
In the limit of geometrical optics we find the Hamiltonian that
determines the trajectories of light rays.
Using a different implementation of the same idea
(transforming Maxwell's equations from the comoving to the
laboratory frame)
Berry and Klein \cite{BK}
have also derived effective scalar and vector potentials
and a Hamiltonian governing light rays and waves.
In addition, we develop a covariant theory of light propagation
in moving media.
In accordance with earlier papers by Gordon \cite{Gordon}
and Pham Mau Quan \cite{PhamMauQuan}
we find that light rays follow zero--geodesic lines measured with 
respect to a certain curved metric in space--time, 
similar to the light propagation in general relativity. 
The metric of the ``glass of water'' depends on the refractive index
and on the flow, and it establishes a fascinating analogy between
moving media and gravitational fields.

Can we see some of the spectacular effects of general relativity
in an earthly laboratory?
Most probably, this would take flow velocities that are comparable 
with the speed of light in the medium.
Recently, dielectrics with incredibly low group velocities 
have been created \cite{Lene}.
These media are far from being as simple as ordinary liquids,
and they are highly dispersive.
For instance, the refractive index reaches unity,
i.e.\ the phase velocity approaches $c$,
at the frequency where the group velocity is lowest,
i.e.\ where the refractive index changes most rapidly.
However, as we show in a separate paper \cite{LPBH},
many phenomena that are conceivable for dispersionless media
find an experimentally feasible analog in dispersive dielectrics.
We can thus employ disperionless media 
as perfectly consistent relativistic models
to understand the key features of some exotic yet realistic effects
of light in moving dielectrics.

The optics of moving media has a long history. 
In 1818 Fresnel \cite{Fresnel} 
discovered theoretically that the speed of light 
$v$ in a uniform yet moving medium of refraction index $n$ 
depends on the medium velocity $u$ as
\begin{equation}\label{fresnel}
v = \frac{c}{n} + \left(1-\frac{1}{n^2}\right)u
\,\,.
\end{equation}
So the effective refraction index $c/v$ is changed when the medium
is moving. 
In 1851 Fizeau \cite{Fizeau}
verified experimentally Fresnel's motional effect (\ref{fresnel}).
In 1895 Lorentz \cite{Lorentz} derived an additional drag effect
that is due to optical dispersion. 
Zeeman \cite{Zeeman} was able to measure Lorentz's effect.
In 1913 Sagnac \cite{Sagnac}
observed phase shifts of light in a rotating interferometer. 
In 1925 Michelson, Gale and Pearson \cite{Michelson}
measured the Sagnac effect of
Earth's motion in an incredible interferometer by 1925 standards. 
And of course, today's fiber gyroscopes prove that 
the interference of light is sensitive to motion.

Despite the long history of optics in moving media,
a sufficiently general theory has been still missing,
with the exception of Berry's and Klein's parallel work \cite{BK}
and of two earlier papers by Pham Mau Quan \cite{PhamMauQuan},
to the best of our knowledge.
In 1908 Minkowski \cite{Minkowski} 
pioneered the modern theory of dielectrics.
In 1923 Gordon published a far--sighted paper \cite{Gordon}
on electromagnetism in dielectrics and in gravitational fields. 
Here he discovered a deep analogy between gravity and dielectric media.
According to Antoci and Mihich \cite{Antoci}, 
Gordon \cite{Gordon} also settled the debate
about Minkowski's \cite{Minkowski} versus Abraham's \cite{Abraham}
energy--momentum tensor in Abraham's favor.
However, Gordon considered only very briefly the geometrical optics 
of moving media.
Pham Mau Quan \cite{PhamMauQuan} studied ray propagation in more detail
but still not exhaustively.
Landau and Lifshitz \cite{LL8} and
Van Bladel \cite{Bladel}
summarize to some extend the current state of the theory,
but do not focus on the propagation of light.
More importantly, 
the motion of the medium has been usually assumed to be uniform.
Exceptions are the papers \cite{Papers} that, however,
treat only special cases such as moving dielectric boundaries
and the consequent modification of Snell's law and the Brewster angle.
Landau and Lifshitz \cite{LL8} write explicitly that they
``neglect slight effects due to the possibility of a velocity
gradient''.

However, these neglected effects are indeed measurable with modern
interferometry (see Sec.\ IV A for an estimation).
Furthermore, effects due to velocity gradients establish
interesting connections between the optics of moving media
and other fields of physics.
Hannay \cite{Hannay} discovered an analogy between light in moving
media and charged matter waves in electromagnetic fields.
The flow ${\bf u}$ turned out to play the role of the 
electromagnetic vector potential.
Hannay used path integrals in paraxial approximation to arrive
at this conclusion.
Cook, Fearn, and Milonni \cite{CFM}
analyzed further the connection between light in moving media and 
charged matter waves, 
assuming relatively slow medium velocities and light that
is perpendicularly polarized with respect to the flow.
The magnetic analogy of light in moving media is particularly
interesting, because the light propagation at a fluid vortex
corresponds to the Aharonov--Bohm effect \cite{AB}
of electron waves that enclose a localized magnetic flux.
Light that travels through a dielectric vortex attains 
an Aharonov--Bohm phase shift. 
On the other hand, atoms that pass an electromagnetic vortex
experience an Aharonov--Bohm effect as well \cite{WilkensChina}.
Interesting quantization effects arise when the atoms
form a macroscopic condensate \cite{LeoPaul}.
The magnetic model of waves in moving media is not restricted to light.
Indeed, Berry {\it et al.} \cite{Berryetal}
report both the theory and an experiment that demonstrates 
an Aharonov--Bohm effect with water waves.
Acoustical analogs of the effect have been observed 
in moving classical media \cite{Roax}
and are predicted for superfluids \cite{Davidowitz}.
However \cite{Gordon},
the magnetic model of light in moving media is only valid
as long as the medium velocities are sufficiently small.
In general, the moving medium turns out to act rather as a curved metric,
i.e. as a gravitational field, on the light.
Note that Unruh \cite{Unruh} arrived at a similar model for
non--relativistic sound in moving fluids that also holds for
superfluids \cite{Russians}.

In Section II we summarize the theory of effects in first order of $u/c$
before we turn to the general case in Sec.\ III.
Section II sets the scene by presenting a short review of partly
published yet not widely known results,
whereas Sec.\ III is the core of our paper.
Here we establish the effective wave equation, 
a Hamiltonian, a Lagrangian and the metric
of light in moving dispersionless media.
Additionally, we show how the magnetic model of 
Sec.\ II and Refs.\ \cite{Hannay,CFM} 
is related to the gravitational one \cite{Gordon,PhamMauQuan}.
Both Sec.\ II and Sec.\ III derive Fresnel's formula (\ref{fresnel}), 
at least to the lowest order in $u/c$,
seen, however, at each case in the light of a distinct physical model.
Quantitative differences between the two concepts are only visible
in higher order.
One example of gravitation--like effects is the light deflection
at a vortex that we analyze in Sec.\ IV and that resembles 
the deflection of light due to Sun's gravity.
In an extreme case the vortex might even appear as an
optical black hole \cite{LPBH}
similar to Unruh's dumb hole \cite{Unruh,Russians}.

\section{Slowly moving media}

Consider a moving nondispersive dielectric medium 
with refractive index $n$
and flow ${\bf u}$.
We allow both $n$ and ${\bf u}$ to vary in space and time.
However, $n$ and ${\bf u}$ shall not change significantly
over the spatial scale of an optical wave length and over one
optical cycle, respectively.
In this section we model light waves by a scalar complex function 
$\psi$, for simplicity.
In particular, we do not consider the polarization of light.
However, we show in Sec.\ III that the propagation of light 
is indeed independent of the polarization,
as long as the medium varies only gradually compared to
optical oscillations.
Furthermore we assume that the medium moves at moderate velocities
such that we can restrict ourselves to effects that occur
within the lowest order in $u/c$.

Our starting point is a simple model:
Imagine that the moving medium consists of small cells or drops.
Each cell shall be small enough such that the refractive index $n$
and the velocity profile ${\bf u}$ of the medium
does not vary significantly.
On the other hand, each cell shall be large compared to the
wave length of light.
We thus assume that in each cell
(in each comoving frame of the medium denoted by primes)
the optical field $\psi$ obeys the wave equation
\begin{equation}
\left(
\nabla'^2 - \frac{n^2}{c^2}\,\frac{\partial^2}{\partial t'^2}
\right)
\psi = 0 
\,\,.
\label{wave1}
\end{equation}
An observer sees the light in the laboratory frame.
To transform the wave equation to the observer's frame,
we write Eq.\ (\ref{wave1}) as
\begin{equation}
\left(
\nabla'^2 - \frac{1}{c^2}\,\frac{\partial^2}{\partial t'^2} - 
\frac{n^2-1}{c^2}\,\frac{\partial^2}{\partial t'^2}
\right)
\psi = 0
\,\,.
\label{wave2}
\end{equation}
We note that the d'Alembert operator 
$\nabla'^2 - \partial^2/(c\partial t')^2$
is a Lorentz invariant,
and thus we transform solely the remaining time derivatives 
in the wave equation (\ref{wave2}).
In the lowest order in $u/c$ a temporal change $\partial/\partial t'$
in the medium frame appears in the laboratory frame 
as the time derivative $\partial/\partial t$ plus the local flow 
${\bf u} \cdot \nabla$.
Therefore, we obtain in first order
\begin{equation}
\left(
\nabla^2 - \frac{n^2}{c^2}\,\frac{\partial^2}{\partial t^2} - 
\frac{n^2-1}{c^2}\,
{\bf u}\cdot\nabla\,\frac{\partial}{\partial t}
\right)
\psi = 0
\,\,.
\label{wave3}
\end{equation}
Note that this derivation of the wave equation for light in slowly
moving media follows Fresnel's original idea \cite{Fresnel,Moeller}
who divided the ether into an invariant part and a second part
that the medium is able to drag.
We will see shortly that Fresnel's formula (\ref{fresnel})
is a direct consequence of the wave equation (\ref{wave3}).

In the limit of Hamilton's geometrical optics 
we represent the optical field $\psi$ in terms of a slowly varying
amplitude and a rapidly changing phase,
\begin{equation}
\psi = {\cal A} e^{iS}
\label{ansatz}
\end{equation}
with
\begin{equation}
S = \int \left( {\bf k} \cdot d{\bf x} - \omega dt \right) 
\,\,.
\label{phase}
\end{equation}
The wave vector ${\bf k}$ corresponds to the momentum of a fictitious
particle that follows a light ray and the frequency $\omega$
plays the role of the Hamiltonian.
We substitute the ansatz (\ref{ansatz}) and (\ref{phase})
into the wave equation (\ref{wave3}),
neglect the variation of the amplitude ${\cal A}$,
and obtain in first order in $u/c$ the dispersion relation
\begin{eqnarray}
0 &=& k^2-\frac{n^2}{c^2}\omega^2 + 
2\omega \frac{n^2-1}{c^2} {\bf u} \cdot {\bf k} 
\label{dispersion1}
\\
&=&
k^2-\frac{n^2}{c^2}
\left(\omega - \frac{n^2-1}{n^2} {\bf u} \cdot {\bf k} \right)^2
\,\,.
\label{dispersion2}
\end{eqnarray}
The Hamiltonian of light rays, $H$, is equal to the frequency $\omega$.
We read immediately from Eq.\ (\ref{dispersion2}) that
\begin{equation}
H = \frac{c}{n} k + \left(1-\frac{1}{n^2}\right) {\bf u}\cdot{\bf k}
\,\,.
\label{h1}
\end{equation}
The ray trajectories are solutions of Hamilton's equations
\begin{equation}
\frac{d{\bf x}}{dt} =  \frac{\partial H}{\partial {\bf k}}
\,,\quad
\frac{d{\bf k}}{dt} = -\frac{\partial H}{\partial {\bf x}}
\,\,.
\label{hamilton}
\end{equation}
The first part of the Hamiltonian, $ck/n$,
describes light rays in a medium at rest
\cite{BornWolf}.
The rays avoid regions of high refractive index
in order to minimize their dimensionless optical path lengths
$\int {\bf k} \cdot d{\bf x} = 
(\omega/c) \int n\, {\bf e_k} \cdot d{\bf x}$
with ${\bf e_k}={\bf k}/k$.
The second part of the Hamiltonian describes Fresnel's 
drag effect.
Indeed, we obtain from Hamilton's equations ({\ref{hamilton})
\begin{equation}
{\bf v} = \frac{d{\bf x}}{dt} =
\frac{c}{n} {\bf e_k} +  \left(1-\frac{1}{n^2}\right){\bf u} 
\,\,, \quad
{\bf e_k} = \frac{\bf k}{k}
\,\,.
\label{fresnelvector}
\end{equation}
This is nothing but the vectorial version of Fresnel's original
formula (\ref{fresnel}).

As has been pointed out earlier \cite{CFM},
a uniform medium in stationary motion acts on light 
in the same way as a magnetic field acts on charged matter waves.
To understand this remarkable connection within our theory of
ray propagation, we introduce a rescaled ray vector ${\bf w}$, or,
equivalently, a reparameterization of the ray trajectory,
\begin{equation}
{\bf w} \equiv k {\bf v} = 
\frac{c}{n} {\bf k} +  \left(1-\frac{1}{n^2}\right)k {\bf u} 
\,\,.
\end{equation}
Let us derive an equation of motion for ${\bf w}$.
First we replace $k{\bf u}$ by $n(\omega/c){\bf u}$
in first order, and get
\begin{equation}
{\bf w} \equiv k {\bf v} = 
\frac{c}{n}\left({\bf k} + \frac{n^2-1}{c^2}\omega{\bf u}\right) 
\,\,.
\label{wk}
\end{equation}
Then we use Hamilton's equations (\ref{hamilton})
and the relation
\begin{equation}
n\frac{\omega}{c}\,\frac{d{\bf u}}{dt} =
n\frac{\omega}{c} ({\bf v} \cdot \nabla){\bf u} = 
({\bf w} \cdot \nabla) {\bf u}
\end{equation}
that is valid in first order.
We obtain the Lorentz--type equation of motion
\begin{equation}
\frac{d{\bf w}}{dt} = \left(1-\frac{1}{n^2}\right)
\left(\nabla \times {\bf u}\right) \times {\bf w}
\,\,.
\label{lorentz}
\end{equation}
Light rays in slowly moving media behave like charged particles
in magnetic fields where the flow ${\bf u}$ appears as a vector
potential. 
The Lorentz--type force (\ref{lorentz}) conserves the modulus
of the modified velocity ${\bf w}$ which is equal to 
$\omega$ in regions where the medium is at rest,
\begin{equation}
w^2=\omega^2
\,\,.
\label{ww}
\end{equation}
We substitute {\bf w} by the right--hand side of Eq.\ (\ref{wk})
and retranslate the resulting dispersion relation into a wave equation,
replacing ${\bf k}\psi$ by $-i\nabla \psi$.
In this way we obtain exactly the Schr\"odinger equation 
of a charged matter wave in a magnetic field \cite{LL3}
\begin{equation}
\left(-i\nabla  + \frac{n^2-1}{c^2}\omega {\bf u}\right)^2\psi =
n^2\frac{\omega^2}{c^2}\psi
\,\,.
\label{schroedinger}
\end{equation}
All these arguments support a magnetic model of light propagation
in moving media \cite{CFM}.
The flow ${\bf u}$ acts as a vector potential that modifies the
relation between the canonical and the kinetic momentum
\begin{equation}
{\bf k} = \frac{n}{c} {\bf w} - \frac{n^2-1}{c^2} \omega {\bf u}
\,\,.
\end{equation}
For example, a rotating rigid glass cylinder will act like a 
homogeneous magnetic field on light that travels inside.
The rotating cylinder will bend light rays,
irrespective of their distance from the rotation axis.
Another example is a vortex that
will act like a thin solenoid, see Sec.\ IV.
Light rays are not bent, but,
similar to the Aharonov--Bohm effect \cite{AB},
rays that enclose the vortex attain a phase difference \cite{CFM}.

\section{Light in moving media}

\subsection{Wave optics}

Let us develop a completely relativistic theory of light propagation
in moving nondispersive media.
Like in Sec.\ II we assume that the refractive index $n$ 
and the flow ${\bf u}$ do not vary significantly over one 
optical wave length and one optical cycle, respectively.
We neglect the dispersion of light, i.e. 
the frequency dependence of $n$.
We allow arbitrary medium velocities and we will employ 
a covariant notation \cite{LL2}.
Our starting point is the following postulate:
In all locally comoving medium frames (denoted by primes)
the electromagnetic field--strength tensor $F'_{\mu\nu}$
shall obey the wave equation
\begin{equation}
\left(
\nabla'^2 - \frac{n^2}{c^2}\,\frac{\partial^2}{\partial t'^2}
\right)
F_{\mu\nu}' = 0 
\,\,.
\label{f1}
\end{equation}
Note that this postulate uses implicitly the assumption that
the refractive index varies only gradually.
Otherwise additional terms become important in the wave equation \cite{LL8},
terms that describe polarization changes 
(at the surfaces of dielectrics, for instance).

Let us transform the wave equation (\ref{f1}) to the laboratory frame.
As a first step we reformulate Eq.\ (\ref{f1}) 
in a covariant notation.
We employ the four--gradients 
\begin{eqnarray}
\partial'_\nu =
\left(\frac{1}{c}\frac{\partial}{\partial t'}, \nabla' \right)
\,\,&,&\quad
\partial'^\nu =
\left(\frac{1}{c}\frac{\partial}{\partial t'}, -\nabla' \right)
\,\,,
\nonumber\\
\partial_\nu = 
\left(\frac{1}{c}\frac{\partial}{\partial t}, \nabla \right)
\,\,&,&\quad
\partial^\nu = 
\left(\frac{1}{c}\frac{\partial}{\partial t}, -\nabla \right)
\,\,,
\end{eqnarray}
and the four--vector field of the medium flow
\begin{equation}
u^\nu =
\gamma \left(1, \frac{{\bf u}}{c}\right)
\,\,,\quad
u_\nu =
\gamma \left(1, -\frac{{\bf u}}{c}\right)
\,\,,
\label{unu}
\end{equation}
with the relativistic factor
\begin{equation}
\gamma= \left(1-\frac{u^2}{c^2}\right)^{-1/2}
\,\,.
\label{gamma}
\end{equation}
In a comoving medium frame the four--vector $u'^\nu$
is locally
\begin{equation}
u'^\nu =
\left(1, {\bf 0}\right)
\,\,.
\end{equation}
Therefore, we can easily write our starting point (\ref{f1})
in the covariant expression
\begin{equation}
\left[
\partial'_\alpha \partial'^\alpha + 
(n^2-1)(u'^\alpha\partial'_\alpha)^2
\right]
F_{\mu\nu}' = 0 
\,\,.
\label{f2}
\end{equation}
Throughout this paper we employ Einstein's summation convention.
When we transform the wave equation (\ref{f2}) 
to the laboratory frame we should transform both the 
derivatives and velocities, and the field--strength tensor.
A Lorentz transformation of a tensor depends of course on the
velocity of the moving frame \cite{LL2}.
Since the medium velocities vary only gradually compared with
the rapid oscillations of $F'_{\mu\nu}$,
we can neglect the derivatives of the Lorentz transformations
of $F'_{\mu\nu}$ in Eq.\ (\ref{f2}).
In other words, the wave equation (\ref{f2}) is valid both
for $F'_{\mu\nu}$ and $F_{\mu\nu}$.
Furthermore, the differential operator
$\partial'_\alpha \partial'^\alpha + (n^2-1)(u'^\alpha\partial'_\alpha)^2$
is a Lorentz scalar,
and therefore we can simply drop the primes in Eq.\ (\ref{f2}),
to arrive at
\begin{equation}
\left[
\partial_\alpha \partial^\alpha + 
(n^2-1)(u^\alpha\partial_\alpha)^2
\right]
F_{\mu\nu} = 0 
\,\,.
\label{f}
\end{equation}
This wave equation describes the propagation of light in a
moving nondispersive medium,
provided that both the refractive index $n$ and the flow ${\bf u}$
varies only gradually.

We see that the final wave equation ({\ref{f}) does not mix 
the components of the field--strength tensor.
The propagation of light beams does not depend on the polarization,
i.e.\ moving media are not birefringent.
This result justifies the scalar model of Sec.\ II.
Note however that the transport of the field amplitudes along
light beams is certainly polarization--dependent.
To describe this effect one should consider,
instead of the wave equation (\ref{f}),
the complete set of Maxwell's equations in moving dielectrics
\cite{Gordon,LL8,Bladel}.

\subsection{Geometrical optics (Hamiltonian)}

How does a moving medium act on light rays?
According to Hamilton's geometrical optics, 
we try the eikonal ansatz \cite{LL2}
\begin{equation}
F_{\mu\nu} = {\cal F}_{\mu\nu}\,e^{iS} + {\rm c.c.}
\end{equation}
with
\begin{equation}
S = \int\left({\bf k} \cdot d{\bf x} - \omega dt \right) =
-\int k_\nu dx^\nu
\,\,.
\label{s}
\end{equation}
Here we have employed the four--differential
\begin{equation}
dx^\nu =
\left(c\, dt, d{\bf x} \right)
\label{dx}
\end{equation}
and the wave four--vector
\begin{equation}
k_\nu =
\left( \frac{\omega}{c}, -{\bf k}\right) =
-\partial_\nu S
\,\,.
\label{k}
\end{equation}
Assuming a rapidly changing phase $S$ compared with the envelope 
${\cal F}_{\mu\nu}$,
we derive from the wave equation (\ref{f}) the
Hamilton--Jacobi equation of light rays,
\begin{equation}
g^{\mu\nu} (\partial_\mu S) (\partial_\nu S) = 0
\,\,,
\label{hj0}
\end{equation}
with 
\begin{equation}
g^{\mu\nu} = 
\eta^{\mu\nu} + (n^2-1) u^\mu u^\nu 
\label{contra}
\,\,,
\end{equation}
using the flat Minkowski metric
\begin{equation}
\eta^{\mu\nu} = \eta_{\mu\nu} =
\left(
\begin{array}{cr}
1 & {\bf 0} \\
{\bf 0} & -{\bf 1}
\end{array}
\right)
\,\,.
\end{equation}
Explicitly, we get the dispersion relation
\begin{equation}
\omega^2 - c^2k^2 + (n^2-1)\gamma^2(\omega - {\bf u \cdot k})^2 = 0
\,\,.
\label{dr}
\end{equation}

To find a Hamiltonian for light beams we solve Eq.\ (\ref{dr})
for $\omega=H$. We obtain
\begin{eqnarray}
H&=&\left(\frac{c^2-u^2}{n^2c^2-u^2}\right)^{1/2}
\left(c^2k^2 - 
\frac{n^2c^2-c^2}{n^2c^2-u^2}({\bf u \cdot k})^2\right)^{1/2} 
\nonumber\\
&& + \,
\frac{n^2c^2-c^2}{n^2c^2-u^2}{\bf u \cdot k}
\,\,.
\label{h}
\end{eqnarray}
In first order in $u/c$ we recognize our previous result (\ref{h1}).
Ray trajectories follow from the Hamiltonian (\ref{h})
as solutions of Hamilton's equations (\ref{hamilton}).
Here the time $t$ plays merely the role of a parameter
to describe the trajectories.
Of course, infinitely many parameterizations exist that 
result in equivalent ray trajectories but stem from
different Hamiltonians.
Does our particular parameterization have a physical meaning?
Let us consider the formal velocity ${\bf v}'$
in the comoving medium frame \cite{Bladel}, 
\begin{equation}
{\bf v}'=
\frac{{\bf v}-{\bf u}+(\gamma-1){\bf u}
({\bf u \cdot v} - u^2)/u^2}
{\gamma\left(1-{\bf u \cdot v}/c^2\right)}
\,\,.
\end{equation}
We use the first of Hamilton's equations (\ref{hamilton})
and the Hamiltonian (\ref{h}), and obtain after some algebra
\begin{equation}
v'^2=\frac{c^2}{n^2}
\,\,.
\label{light}
\end{equation}
Light travels with the velocity of light.
Therefore, we can identify the formal ray parameter $t$
with the physical travel time of light in the laboratory frame.

Apart from elucidating the physical meaning of time
for our Hamiltonian,
the relation (\ref{light}) is simply an explicit conservation law
during the ray propagation.
Conservation laws are connected to symmetries.
In fact, the Hamiltonian (\ref{h}) has the remarkable structure
\begin{equation}
H = k\, h(\zeta)
\,\,,\quad
\zeta={\bf u}\cdot{\bf e_k}
\,\,,\quad
{\bf e_k} = \frac{{\bf k}}{k}
\,\,.
\label{hstructure}
\end{equation}
Consequently,
\begin{equation}
{\bf v} = \frac{\partial H}{\partial {\bf k}} =
h\,{\bf e_k} + 
\frac{\partial h}{\partial \zeta}\, 
{\bf e_k} \times ({\bf u} \times {\bf e_k})
\,\,.
\label{vstructure}
\end{equation}
We see that the velocity vector ${\bf v}$
is independent of the wave number $k$.
The trajectories of light beams in moving media 
do not depend on the wave properties of light.
Geometrical optics involves indeed solely the geometrical aspects
of light rays.

\subsection{Lagrangian}

We have studied the propagation of light in moving media 
in the spirit of Hamilton's geometrical optics.
Can we find a Lagrangian as well?
Let us calculate the Lagrangian directly from the Hamiltonian,
using the structure (\ref{hstructure}) and (\ref{vstructure})
of the Hamiltonian and the velocity vector, respectively.
We get
\begin{equation}
L={\bf k}\cdot{\frac{\partial H}{\partial {\bf k}}} - H = 0
\,\,.
\end{equation}
The Lagrangian vanishes.
Furthermore,
we cannot express the canonical momentum ${\bf k}$
in terms of the velocity ${\bf v}$,
because, according to Eq.\ (\ref{vstructure}),
${\bf v}$ does not depend on the modulus of ${\bf k}$.
The encountered problems in introducing a Lagrangian 
for light rays in moving media
have been known for the special case of light in vacuum \cite{LL2}
and have been attributed \cite{LL2}
to the vanishing rest mass of light.

Let us provide light with an artificial rest mass $m$
that we let approach zero at a later, appropriate moment.
We replace the right--hand side of the wave equation (\ref{f})
by $m^2 c^2 F_{\mu\nu}$,
and we obtain in the limit of geometrical optics the
Hamilton--Jacobi equation
\begin{equation}
g^{\mu\nu} (\partial_\mu S) (\partial_\nu S) = m^2c^2
\,\,.
\label{hj}
\end{equation}
We wish to find a Lagrangian representation of the phase $S$ such that
\begin{equation}
S = \int L\, dt
\,\,.
\end{equation}
For this we introduce the matrix
\begin{equation}
g_{\mu\nu} = 
\eta_{\mu\nu} + \left(\frac{1}{n^2}-1\right) u_\mu u_\nu 
\label{co}
\,\,.
\end{equation}
One verifies easily that $g_{\mu\nu}$ is the inverse of of $g^{\mu\nu}$,
utilizing that $u_\nu u^\nu$ is normalized to unity, 
\begin{equation}
g_{\mu\alpha}g^{\alpha\nu} = \delta_\mu^\nu
\,\,.
\label{inverse}
\end{equation}
We obtain the solution of the Hamilton--Jacobi equation
\begin{equation}
S = -mc \int ds 
\label{ss}
\end{equation}
with 
\begin{equation}
ds^2 = g_{\mu\nu} dx^\mu dx^\nu
\,\,.
\end{equation}
To prove this result we note that the wave four--vector gives 
\begin{equation}
k_\mu = -\partial_\mu S = mc\, g_{\mu\nu} \frac{dx^\nu}{ds}
\,\,, 
\label{k_mu}
\end{equation}
and thus satisfies the Hamilton--Jacobi equation (\ref{hj})
\begin{equation}
g^{\mu\nu} k_\mu k_\nu = 
m^2 c^2 g_{\mu\nu} \frac{dx^\mu}{ds} \, \frac{dx^\nu}{ds} =
m^2 c^2
\,\,.
\end{equation}
The expression (\ref{ss}) of the phase $S$ has the desired
Lagrangian structure, 
\begin{equation}
S = -mc \int \sqrt{g_{\mu\nu} \frac{dx^\mu}{dt} \,
  \frac{dx^\nu}{dt}}\,dt =
\int L\, dt
\,\,,
\end{equation}
with the explicit Lagrangian
\begin{equation}
L = -mc\, \sqrt{c^2 - v^2 + \left(\frac{1}{n^2}-1\right)
\gamma^2\left(c-\frac{{\bf u} \cdot {\bf v}}{c}\right)^2}
\,\,.
\end{equation}

\subsection{Metric}

According to the action principle, light rays minimize the phase $S$.
Because the phase ({\ref{ss}) is proportional to the artificial mass $m$,
the minima of $S$ do not depend on the numerical value of $m$ at all.
We thus arrive at an entirely geometrical picture of
ray optics in moving media,
in complete analogy to the equivalence principle
of general relativity.
Light rays are geodesic lines with respect to Gordon's metric \cite{Gordon}
\begin{eqnarray}
ds^2
&=&
\eta_{\mu\nu} dx^\mu dx^\nu + \left(\frac{1}{n^2}-1\right)
\left(u_\mu dx^\mu\right)^2
\label{metric}
\\
&=&
\eta_{\mu\nu} dx'^\mu dx'^\nu + \left(\frac{1}{n^2}-1\right) c^2dt'^2
\nonumber\\
&=&
\frac{c^2}{n^2} dt'^2-d{\bf x}'^2 
\,\,.
\end{eqnarray}
While traveling in a moving medium,
light minimizes the proper time measured with respect to $c/n$.
And, because light actually travels with the velocity of light 
(\ref{light}), 
the proper time is zero on the ray trajectory.
Light rays follow zero--geodesic lines of the metric
(\ref{metric}).

A moving medium acts as a curved space--time on light.
In terms of Riemann's geometry,
a medium with uniform refractive index $n$ turns out to have 
the scalar curvature \cite{LL2}
\begin{eqnarray}
R & = &
\frac{(n^2-1)^2}{2n^2}
(u^\mu u^\nu-\eta^{\mu\nu})
(\partial_\mu u^\iota) (\partial_\nu u_\iota)
\nonumber\\
&& + \,
\frac{1-n^4}{2n^2} 
(\partial_\mu u^\nu) (\partial_\nu u^\mu) 
\nonumber\\
&& + \,
(1-n^2)\left(
2u^\nu\partial_\nu\partial_\mu u^\mu + (\partial_\mu u^\mu)^2
\right)
\,\,,
\end{eqnarray}
as one obtains after some algebra.
When the flow obeys the continuity equation,
\begin{equation}
\partial_\nu u^\nu = 0
\,\,,
\label{continuation}
\end{equation}
the curvature simplifies to 
\begin{eqnarray}
R & = &
\frac{(n^2-1)^2}{2n^2}
(u^\mu u^\nu-\eta^{\mu\nu})
(\partial_\mu u^\iota) (\partial_\nu u_\iota)
\nonumber\\
&& + \,
\frac{1-n^4}{2n^2} 
(\partial_\mu u^\nu) (\partial_\nu u^\mu) 
\,\,.
\end{eqnarray}
In curved space--time the distinction between co- and 
contravariant objects is particularly profound.
Let us introduce the contravariant wave vector
with respect to the metric of the moving medium,
\begin{equation}
k^\mu \equiv g^{\mu\nu} k_\nu
\,\,.
\label{kcontra}
\end{equation}
According to Eqs.\ (\ref{inverse}) and (\ref{k_mu})
the $k^\mu$ four--vector is proportional to the four--velocity,
\begin{equation}
k^\mu = mc \, \frac{dx^\mu}{ds} = mc\,\frac{dt}{ds}\,\frac{dx^\mu}{dt}
\,\,.
\end{equation}
Because $dx^0/dt$ is equal to $c$, by definition (\ref{dx}),
we obtain
\begin{equation}
k^\mu = \frac{k^0}{c}\,\frac{dx^\mu}{dt}
\,\,,
\end{equation}
a relation that remains valid for vanishing $m$ and $ds$.
In the magnetic model of light propagation in moving media, 
see Sec.\ II,
the three--dimensional ${\bf k}$ vector differs from the velocity
by an additional term that is proportional to the flow
and that plays the role of a vector potential.
However, as we have seen, the contravariant $k^\mu$ vector
is in fact proportional to the velocity $dx^\mu/dt$.
Therefore, the appearance of the flow as a vector potential
is an illusion that we get when we do not discriminate
between co-- and contravariant vectors,
without yet appreciating the inherent space--time geometry of
geometrical optics of moving media.

\subsection{Slowly moving media}

In Sec.\ II we summarized the magnetic model \cite{Hannay,CFM}
of light in slowly moving media.
Let us formulate the first--order effects in terms of the
geometrical model \cite{Gordon}.
To the lowest order in $u/c$ we obtain the metric
\begin{equation}
ds^2 = \frac{1}{n^2}\,(c dt)^2 - d{\bf x}^2 +
2\,\frac{n^2-1}{n^2}\,\frac{\bf u}{c} \cdot d{\bf x}\, c dt
\,\,.
\end{equation}
Therefore, the covariant metrical tensor $g_{\mu\nu}$ 
is given by the matrix
\begin{equation}
g_{\mu\nu} =
\left(
\begin{array}{cc}
{\displaystyle \frac{1}{n^2}} & 
{\displaystyle \frac{n^2-1}{n^2}\,\frac{\bf u}{c}} \\
{\displaystyle \frac{n^2-1}{n^2}\,\frac{\bf u}{c}} & 
{\displaystyle -{\bf 1}}
\end{array}
\right)
\,\,.
\end{equation}
To find the contravariant metrical tensor we compare
the dispersion relation (\ref{dispersion1})
with Eqs.\ (\ref{k}) and (\ref{hj0}),
and read off the result
\begin{equation}
g^{\mu\nu} =
\left(
\begin{array}{cc}
{\displaystyle n^2} & 
{\displaystyle (n^2-1)\frac{\bf u}{c}} \\
{\displaystyle (n^2-1)\frac{\bf u}{c}} & 
{\displaystyle -{\bf 1}}
\end{array}
\right)
\,\,.
\end{equation}
In the limit of low medium velocities,
three--dimensional space 
appears to be flat to light that travels in the medium,
yet four--dimensional space--time is curved.
Let us compare co- and contravariant wave vectors.
As a consequence of Eqs.\ (\ref{wk}) and (\ref{k}) 
and of the definition (\ref{kcontra}),
the space components $k^i$ of the contravariant 
$k^\mu$ are given by
\begin{equation}
k^i = {\bf k} + (n^2-1)\,\frac{\omega}{c}\,{\bf u} =
\frac{n}{c}\,{\bf w} = \frac{n}{c}\, k\, {\bf v}
\,\,.
\label{final}
\end{equation}
Therefore, the contravariant wave vector 
appears as the kinetic momentum,
in contrast to the canonical one that is represented by
the covariant wave vector.
Finally,
we solve Eq.\ (\ref{final}) for ${\bf k}/k$
and arrive at the vectorial form (\ref{fresnelvector}) of
Fresnel's classic result (\ref{fresnel}),
seen here in the Riemann--geometrical model of light in moving media.

\section{Light around a vortex}

\subsection{Optical Aharonov--Bohm effect}

Consider a vortex flow in an incompressible liquid.
The vortex may be created by the action of a mixer or simply by
letting the liquid flow out off the container.
To a good approximation the velocity profile of the vortex is given,
using cylindrical coordinates, by the expression 
\cite{Lamb}
\begin{equation}
{\bf u} = \frac{\cal W}{r}\,{\bf e}_\varphi
\,\,.
\label{vortex1}
\end{equation}
Let light travel through the whirling liquid.
In first order of $u/c$, 
light experiences the medium flow in the same way 
as a charged matter wave experiences a vector potential 
\cite{Hannay,CFM}.
In particular, the vortex (\ref{vortex1}) corresponds to
an infinitely thin solenoid that generates a strong magnetic
field inside yet no field outside,
\begin{equation}
\nabla \times {\bf u} = {\bf 0}
\,\,.
\end{equation}
However, as Aharonov and Bohm discovered in their seminal 
1959 paper \cite{AB}, a charged matter wave 
will attain a phase shift without feeling a force,
and so will light.
We compare the Schr\"odinger equation of Aharonov and Bohm \cite{AB}
with the wave equation (\ref{schroedinger})
and read off the optical Aharonov--Bohm phase shift
\begin{equation}
\varphi_{_{AB}} = 2\pi \nu_{_{AB}}
\,\,,\quad
\nu_{_{AB}} = \frac{\omega}{c} (n^2-1) \frac{\cal W}{c}
\,\,.
\label{abphase}
\end{equation}
How large is the effect?
For estimating the order of magnitude we assume a refractive index 
$n$ of $1.5$ and an optical frequency $\omega$ of 
$3\times 10^{15}{\rm s}^{-1}$.
Furthermore we assume that the liquid circulates at a radius $r$
of $2\,{\rm cm}$ with a frequency $u/r$ of $10\, {\rm s}^{-1}$.
We obtain a phase shift $\varphi_{_{AB}}$ of $10^{-3}$.
The effect is small, yet one could enhance it significantly
by letting the light travel through the liquid many times
in an interferometer.
In this case the phase shift $\varphi_{_{AB}}$ is effectively 
multiplied by the number of round--trips.
One could also perceive the interferometer as a resonator 
where the effect of the moving medium changes the spatial mode profile.
The precision of modern interferometry is certainly sufficient 
to detect the optical Aharonov--Bohm effect.
(Remember that Fizeau \cite{Fizeau} 
has seen the precursor of the effect as early as 1851.)
Figure 1 illustrates the long--range behavior of light waves
that pass a vortex flow.
\begin{figure}[h]
  \begin{center}
    \includegraphics[width=40pc]{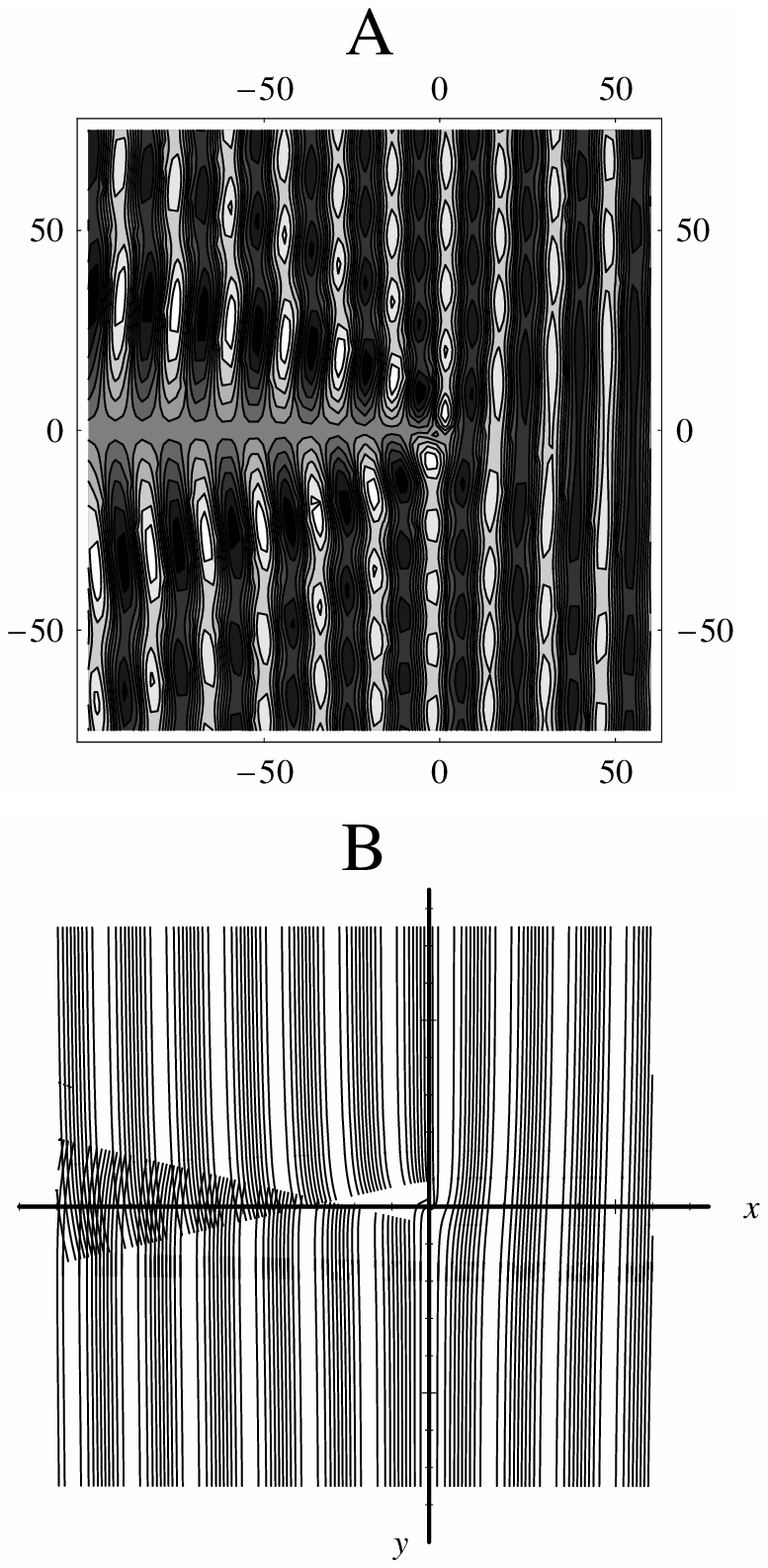}
    \vspace*{5mm}

    \caption{Light wave passing through a vortex flow.
    Picture (A) illustrates the optical Aharonov--Bohm effect.
    Light is incident from the right, and 
    we plot the real part of the optical field $\psi$
    that is subject to the wave equation (\ref{schroedinger}).
    We use for $\psi$ the well--known Bessel series 
    due to Aharonov and Bohm [20].
    Picture (B) sketches the wave patterns of light,
    taken into account the slight ray bending that we discuss 
    in Sec.\ IV C. We use the same scale as in (A).
    Before reaching the vortex the wave fronts bend and then
    they split at the vortex core.
    Behind the core the two parts of the wave interfere.
    We created the picture by following a bundle of exact ray
    trajectories (\ref{rels}).
    A wave front is defined by the points reached at a particular
    value of the time parameter.
    Although this approach does not provide information about
    the wave amplitudes, the picture is quite accurate, because
    mostly only one part of the split wave contributes to interference,
    except in regions very close to the cut at the negative $x$ axis.
    We see that picture (B) reproduces the prominent long--range
    features of the Aharonov--Bohm wave depicted in (A).}
    \label{fig:figure1}
  \end{center}
\end{figure}

\subsection{Relativistic vortex}

Let us analyze vortex effects to higher order in $u/c$.
Note however that the non--relativistic vortex (\ref{vortex1})
permits medium velocities that exceed the speed of light,
when taken seriously near the vortex core where higher--order
effects are to be expected.
Let us therefore seek a proper relativistic vortex.
We anticipate that the flow is still circular,
\begin{equation}
{\bf u} = u(r) {\bf e}_\varphi
\,\,,
\label{ur}
\end{equation}
but that we must correct the non--relativistic velocity profile
(\ref{vortex1}) by the relativistic $\gamma$ factor (\ref{gamma}),
\begin{equation}
u = \frac{\cal W}{\gamma r}
\,\,,
\label{uw}
\end{equation}
to prevent the medium from becoming tachyonic.
Given the velocity profile ({\ref{uw}), 
we solve Eq.\ (\ref{gamma}) for $\gamma$, and obtain
\begin{equation}
\gamma = 
\sqrt{1+\frac{{\cal W}^2}{c^2 r^2}}
\,\,.
\label{gammavortex}
\end{equation}
Therefore, the four--flow (\ref{unu}) is finally
\begin{equation}
u^\nu = 
\left(
\sqrt{1+\frac{{\cal W}^2}{c^2 r^2}}, 
\frac{\cal W}{cr}\, {\bf e}_\varphi \right)
\,\,.
\label{vortex}
\end{equation}
Of course, so far we have simply guessed the proper relativistic
generalization of the vortex flow (\ref{vortex1}).
To prove that our result is indeed physically meaningful,
we shall show that the four--flow (\ref{vortex})
is consistent with relativistic hydrodynamics \cite{LL6}.
First we note that any circular flow (\ref{ur})
satisfies the equation of continuation (\ref{continuation}).
Then we prove that the particular four--flow (\ref{vortex})
is a solution of the relativistic Euler equation.
For this we construct the energy--momentum tensor \cite{LL6}
\begin{equation}
T^{\mu\nu} = (\rho c^2 + p) u^\mu u^\nu - p\, \eta^{\mu\nu}
\end{equation}
and assume the pressure profile
\begin{equation}
p = \rho c^2 \exp\left(-\frac{{\cal W}^2}{2c^2r^2}\right) - 
\rho c^2
\,\,.
\label{pressure}
\end{equation}
Here $\rho$ denotes the constant mass density of the 
incompressible fluid.
We obtain by direct calculation that $T^{\mu\nu}$
satisfies the Euler equation 
\begin{equation}
\partial_\mu T^{\mu\nu} = 0
\,\,.
\end{equation}
This proves that the vortex (\ref{vortex}) is indeed 
a possible flow that, as we have seen as well, 
generates the pressure profile (\ref{pressure}).
At sufficiently large distances from the vortex core the 
pressure behaves like
\begin{equation}
p \sim - \frac{\rho{\cal W}^2}{2r^2}
\,\,.
\end{equation}
The rapidly falling pressure will attract the surface of the liquid
and create a hole at the vortex core.
Let us nevertheless treat the hole as being infinitely thin.
Near the vortex core the pressure is finite
\begin{equation}
\lim_{r\rightarrow 0} p = -\rho c^2
\,\,,
\end{equation}
and the medium velocity reaches the speed of light
\begin{equation}
\lim_{r\rightarrow 0} u = c
\,\,.
\end{equation}
At large distances from the core the medium flows as described
by the non--relativistic Eq.\ (\ref{vortex1}).
Of course, dramatically relativistic vortices are not to be
expected to be created in experiments on Earth in the near future, 
but they might exist as astronomical objects.

\subsection{Light bending}

Let us study the propagation of light rays around the relativistic
vortex (\ref{vortex}).
For this we solve the Hamilton--Jacobi equation (\ref{hj})
that reads explicitly
\begin{equation}
{\dot S}^2 - c^2(\nabla S)^2 +
(n^2-1)\gamma^2\, (\dot S + {\bf u}\nabla S)^2
= 0
\,\,.
\label{hjexplicit}
\end{equation}
We obtain for any circular flow (\ref{ur}) the solution
\begin{equation}
S=\frac{\omega}{c} \left[-c t + l\varphi + R(r)\right]
\label{rels}
\end{equation}
with
\begin{equation}
\left(\frac{dR}{dr}\right)^2 =
1 - \frac{l^2}{r^2} + 
\frac{n^2-1}{\displaystyle 1-\frac{u^2}{c^2}}
\left(1-\frac{ul}{cr}\right)^2
\,\,.
\label{relr}
\end{equation}
Let us first study light rays that travel sufficiently far
outside the vortex core.
We use our velocity profile (\ref{uw}) with the factor 
(\ref{gammavortex}),
expand $(dR/dr)^2$ to quadratic order in $r^{-1}$, and get
\begin{equation}
\left(\frac{dR}{dr}\right)^2 =
n^2 - \frac{l_{_{AB}}^2}{r^2} + 
(n^2-1) \frac{n^2{\cal W}^2}{c^2r^2}
\end{equation}
with
\begin{equation}
l_{_{AB}} = l + (n^2-1) \frac{\cal W}{c}
\,\,.
\end{equation}
As we shall see below, $l_{_{AB}}$ plays the role of the kinetic
angular momentum that is altered by the optical Aharonov--Bohm effect.
According to canonical mechanics \cite{LL1}
trajectories are given requiring that the derivative
of the action with respect to generalized canonical momenta 
is a set of constants.
In particular we put
\begin{equation}
\frac{\partial S}{\partial l} = \frac{\omega}{c} \varphi_0
\,\,.
\end{equation}
In this way we obtain the trajectories
\begin{equation}
\varphi - \varphi_0 =
\int_{r_0}^r 
\frac{l_{_{Ab}}\, r^{-2}\, dr}
{\displaystyle \sqrt{n^2 - \frac{l_{_{AB}}^2}{r^2} + 
(n^2-1)\frac{n^2{\cal W}^2}{c^2r^2}}}
\end{equation}
with
\begin{equation}
r_0^2 = \frac{l_{_{AB}}^2}{n^2} - (n^2-1)\,\frac{{\cal W}^2}{c^2} 
\,\,,
\label{r0}
\end{equation}
and, after solving the integral,
\begin{equation}
\varphi - \varphi_0 = \frac{l_{_{AB}}}{n r_0} 
{\rm arccos} \left(\frac{r_0}{r}\right)
\,\,.
\label{phir}
\end{equation}
To first order in $u/c$ we disregard the ${\cal W}^2/c^2$ term
in Eq.\ (\ref{r0}) and get from Eq.\ (\ref{phir})
\begin{equation}
r \cos(\varphi - \varphi_0) = r_0
\,\,.
\end{equation}
Therefore,
far outside the vortex core light travels along straight lines
(in accordance with our previous considerations on the
optical Aharonov--Bohm effect).
Let us study indications of higher--order effects.
At infinity the angle $\varphi$ approaches 
\begin{equation}
\varphi_\infty \equiv \lim_{r\rightarrow\infty} \varphi =
\varphi_0 + \frac{\pi}{2}\,\frac{l_{_{AB}}}{n r_0}
\,\,.
\end{equation}
The impact parameter of the ray is given by
\begin{equation}
b \equiv \lim_{r\rightarrow\infty} 
r \sin(\varphi-\varphi_\infty) =
-\frac{l_{_{AB}}}{n}
\,\,.
\end{equation}
This result shows explicitly that $l_{_{AB}}$ corresponds
indeed to the kinetic angular momentum.
When the light leaves the vortex region the ray is deflected
by the angle \cite{LL1}
\begin{eqnarray}
\varphi_{_D} &=& \pi - 2(\varphi_\infty - \varphi_0)
= \pi - \pi \frac{l_{_{AB}}}{n r_0}
\nonumber\\
&\approx&
\frac{\pi}{2} \frac{(n^2-1){\cal W}^2}{b^2c^2} 
\end{eqnarray}
to second order in ${\cal W}/c$.
Similar to the bending of star light due to Sun's gravity \cite{LL2},
light is attracted by a vortex and is consequently slightly
deflected.
We express $\varphi_{_D}$ in terms of the
first--order phase shift (\ref{abphase}),
and obtain to second order in ${\cal W}/c$
\begin{equation}
\varphi_{_D} = \frac{n^2}{8\pi(n^2-1)}\,\nu_{_{AB}}^2\,
\frac{\lambda^2}{b^2}
\,\,,\quad
\lambda = \frac{2\pi c}{\omega n}
\,\,.
\end{equation}
The deflection angle is extremely small for small
Aharonov--Bohm phases $2\pi\nu_{_{AB}}$. 
Figure 2 illustrates the bending of light at a vortex.
The figure shows also light rays in the immediate vicinity of the core,
a case that we shall consider in the next subsection.
\begin{figure}[htbp]
  \begin{center}
    \includegraphics[width=20.5pc]{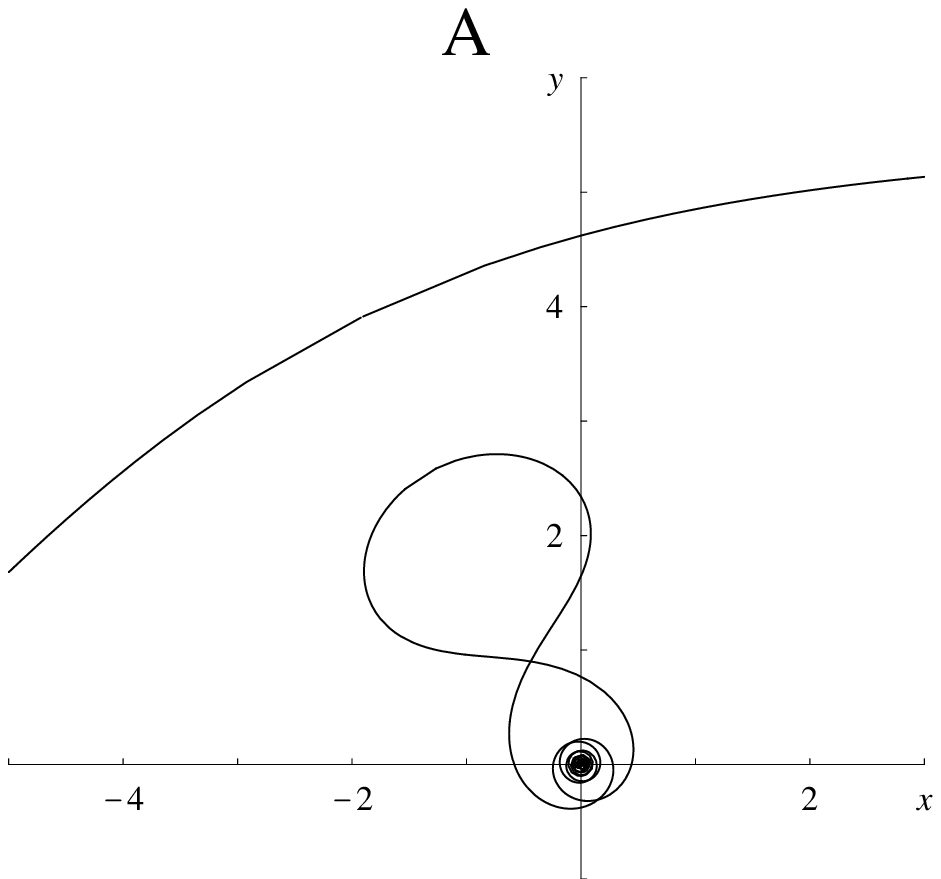}\\
    \includegraphics[width=20.5pc]{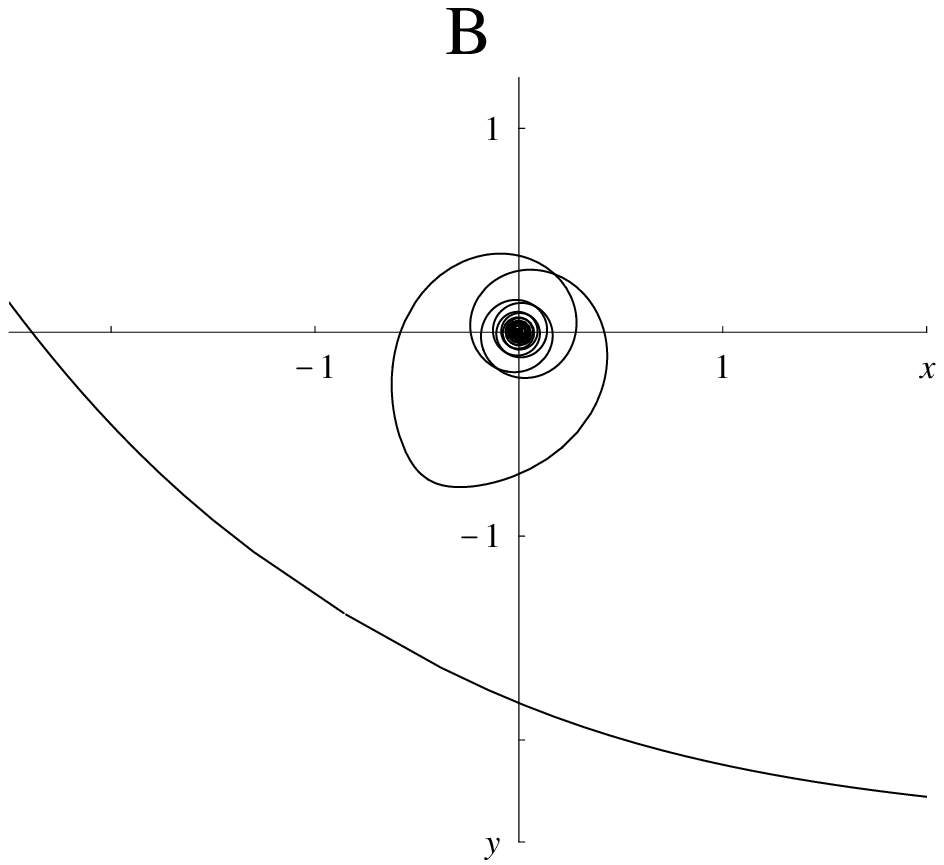}
    \vspace*{2mm}

    \caption{Trajectories of light rays in the vicinity of a vortex flow.
    Each picture shows a pair of trajectories belonging to the same
    value of the angular momentum.
    In one case the light ray approaches the vortex from infinity,
    is bent, and disappears to infinity.
    In the other case the light stems from the origin and falls back
    after having reached its maximal range from the vortex core.
    At distances not covered by these trajectories no light rays are
    allowed that belong to the chosen 
    particular value of the angular momentum.
    (A) Negative angular momentum (positive impact parameter).
    Light swims against the current.
    (B) Positive angular momentum (negative impact parameter).
    Light travels with the flow.
    }
    \label{fig:figure2}
  \end{center}
\end{figure}

\subsection{Optical black hole}

A vortex attracts light like any other test particle,
because of the rapidly falling pressure profile (\ref{pressure}).
Can light fall into the vortex core?
To answer this question we study the turning points of 
light rays, using the complete relativistic action (\ref{rels})
with the radial component (\ref{relr}).
Turning points are the zeros of $(dR/dr)^2$ because beyond
these points the action $S$ is purely imaginary and hence
covers a forbidden region.
We represent Eq.\ (\ref{relr}) as
\begin{equation}
\left(\frac{dR}{dr}\right)^2 =
\gamma^2 
\left[n^2\left(1 - \frac{ul}{cr} \right)^2 - 
\left(\frac{u}{c} - \frac{l}{r} \right)^2\right]
\,\,.
\end{equation}
We see that each radius $r$ can be reached as a turning point
of two trajectories that are characterized by the angular momenta
\begin{equation}
l_\pm = r\,\frac{nc\pm u}{nu\pm c}
\,\,.
\end{equation}
The point may be an inner or an outer turning point,
depending on the sign of
\begin{equation}
\zeta_\pm \equiv 
\left.
\frac{\partial}{\partial r}
\left(\frac{dR}{dr}\right)^2 \right|_{l_\pm}
\,\,.
\end{equation}
The vortex may emit light that begins to fall back to the core
at an inner turning point with negative $\zeta$.
Or, alternatively, incident light from outside the vortex is bent
and comes to the closest distance to the core at an outer
turning point with positive $\zeta$.
To distinguish inner and outer turning points we calculate 
$\zeta_\pm$ and write the resulting expression 
in terms of the derivative of $l_\pm$ regarded as a function of $r$.
We obtain 
\begin{equation}
\zeta_\pm =
\pm 2n \,\frac{dl_\pm}{dr}
\,\,,
\quad
\frac{dl_\pm}{dr} =
\frac{l_\pm}{r} - \frac{n^2-1}{(nu\pm c)^2}\,ru'c
\,\,.
\end{equation}
The velocity profile $u(r)$ of the vortex (\ref{vortex})
is monotonically decreasing and positive.
(We assume a positive ${\cal W}$.
Otherwise one could invert the system of coordinates
to arrive at a positive vorticity.)
Therefore, $l_+$ is positive and also yields a positive $\zeta_+$.
Given a positive angular momentum,
any point can be reached as an outer turning point of incident light rays.
Let us turn to the $l_-$.
The angular momentum $l_-$ of a turning point is negative 
outside the radius where the medium velocity reaches 
the speed of light in the medium, $c/n$. 
Inside this radius the trajectory that corresponds to $l_-$ 
approaches an inner turning point, because $\zeta_-$ is negative.
On the other hand, for negative angular momenta a critical
radius $r_s$ exists where $\zeta_-$ vanishes.
Beyond this radius all light rays with negative $l$ 
must fall into the vortex core.
The vortex appears as a black hole to light with negative
angular momenta.
Let us calculate the optical Schwarzschild radius $r_s$.
We utilize that 
\begin{equation}
ru' = -u \left(1-\frac{u^2}{c^2}\right)
\end{equation}
for the relativistic vortex (\ref{vortex}) and arrive
at a third--order equation for the velocity $u_s$
at the Schwarzschild radius where $dl_-/dr$ vanishes,
\begin{equation}
(n^2-1) \xi^3 + n \xi^2 - 2n^2 \xi + n = 0
\,\,,\quad 
\xi=\frac{u_s}{c}
\,\,.
\end{equation}
This equation has three real solutions labeled by $i \in \{1, 2, 3\}$,
\begin{eqnarray}
u_i(n)&=&\frac{2 nc\sqrt{6n^2-5}\,\cos
  \left[\alpha_i(n)\right]}{3(n^2-1)}-\frac{nc}{3(n^2-1)}
\nonumber\\
\alpha_i(n) &=&
\frac{1}{3}\left[2\pi (i-1)+ \beta(n)\right]
\,\,,
\nonumber\\
\beta(n) &=&
\arccos\left(-\frac{27-70n^2+45n^4}{2n^2(6n^2-5)^{3/2}}\right)
\,\,.
\label{ui}
\end{eqnarray}
The values of the physical solution must lie in the interval  
$0\leq u_i <c  $ in order to be velocities allowed by the theory of
relativity. 
The refractive index $n$ is confined to the interval $1\leq n < \infty$.
The only function assuming physically permitted
values at the ends of this interval is $u_3(n)$ with
\begin{equation}
u_3(1)=c
\,\,,\quad
\lim_{n \rightarrow \infty} u_3(n)=0
\,\,.
\end{equation}
In order to prove that the results for intermediate values of $n$ 
are physical as well, we show that $u_3(n)$ 
is a monotonically decreasing function. 
The first term in (\ref{ui}) is a
product of two positive and monotonically decreasing functions 
and thus is a positive monotonically decreasing function itself. 
So is the second term that is subtracted. 
In order to show that the total expression
stays positive one can employ the estimates
\begin{eqnarray}
\arccos(x)& \geq & -\frac{1}{2} \sqrt{1+x}\,\pi+\pi
\,\,, \quad \mbox{for }-1\leq x \leq 0
\,\,,
\nonumber\\
\cos (x) & \geq & \frac{3}{\pi}x-\frac{9}{2}, 
\quad \mbox{for }\frac{\pi}{2}\leq x\leq \pi.
\end{eqnarray}
This leads to an algebraic inequality that can be solved in a
standard way, thus proving our assumption. 
In an analogous way one can show that 
$u_1$ and $u_2$ do not assume physical values.
Finally, to calculate the optical Schwarzschild radius $r_s$,
we solve Eqs.\ (\ref{uw}) and (\ref{gammavortex}) for $r$, and get
\begin{equation}
r_s = \frac{{\cal W}}{u_3} \sqrt{1-\frac{u_3^2}{c^2}}
\,\,.
\end{equation}
Optical black holes could be created using highly dispersive 
dielectrics \cite{Lene,LPBH}.
These media are distinguished by an extremely low group velocity of
light.
To some extend, they resemble dispersionless dielectrics with
an extremely large refractive index.
Let us therefore consider the limit of large $n$.
We obtain from Eq.\ (\ref{ui}) the critical flow velocity
\begin{equation}
u_s = u_3 \sim \frac{c}{2n}
\,\,.
\end{equation}
A light ray with a negative angular momentum,
i.e. light that swims against the current,
is trapped when the flow velocity approaches half of the speed 
of light in the medium.
Figure 3 illustrates the fall into the optical black hole.
\begin{figure}[htbp]
  \begin{center}
    \includegraphics[width=20.5pc]{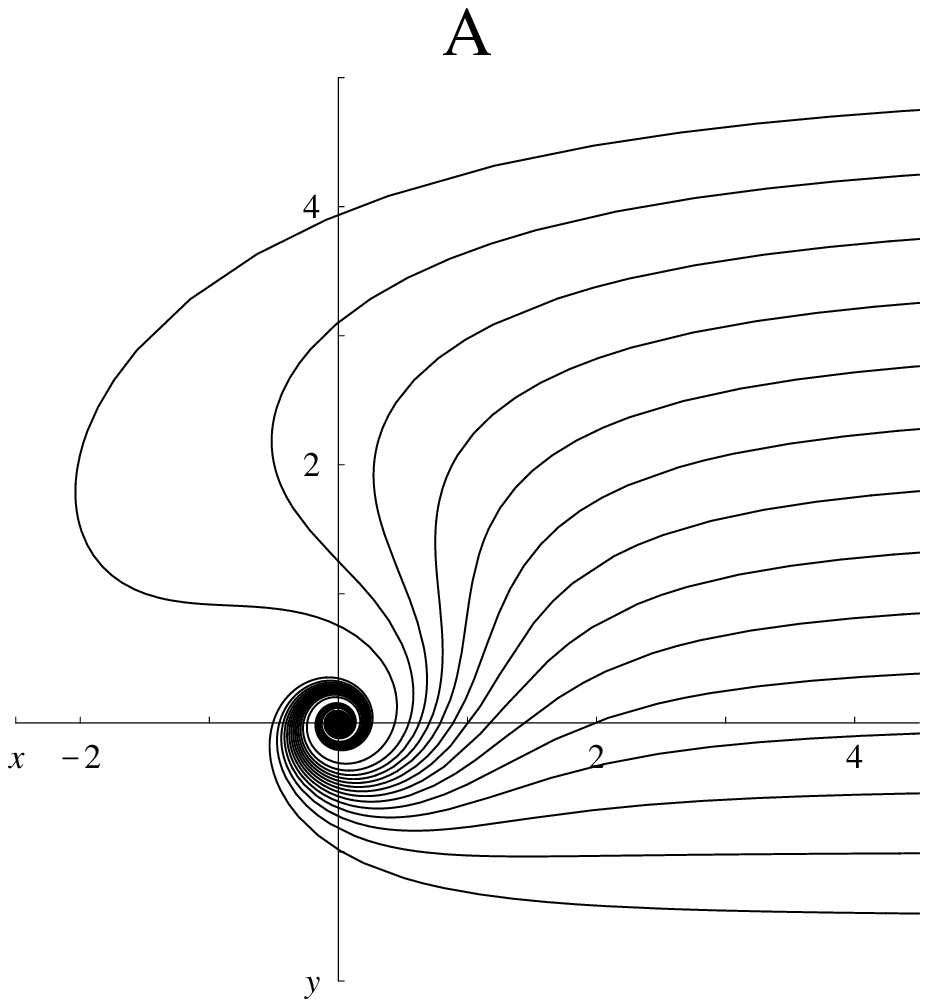}\\
    \includegraphics[width=20.5pc]{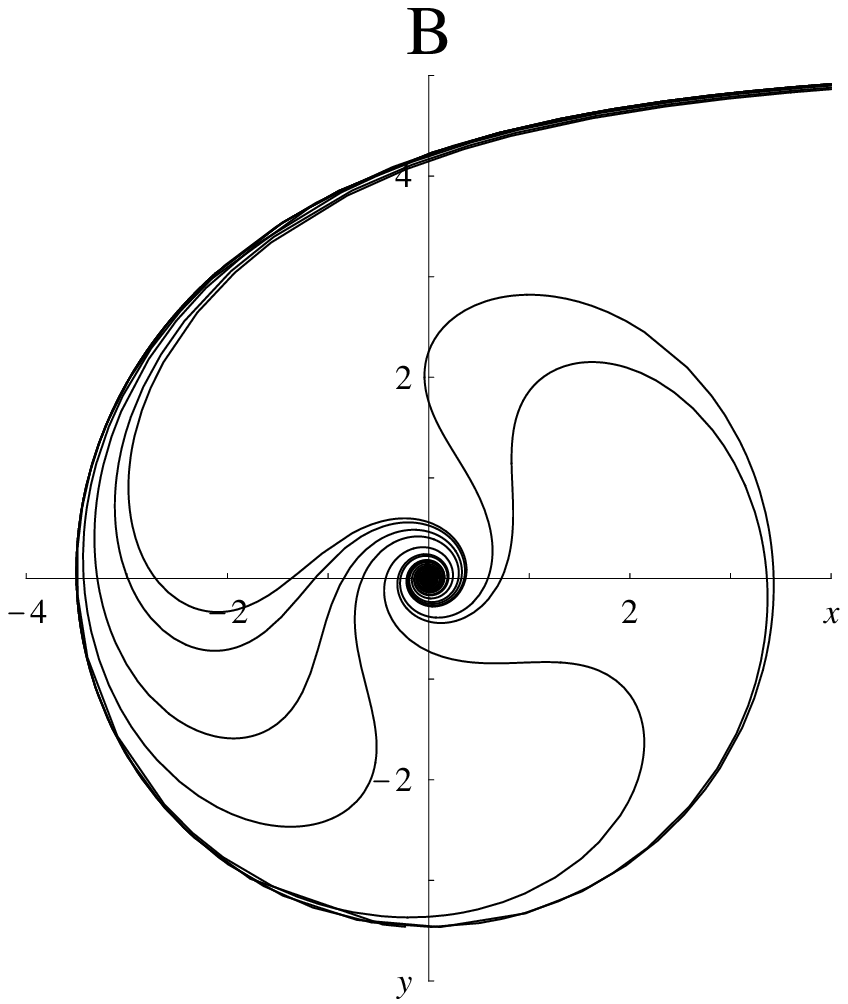}
    \vspace*{2mm}

    \caption{Optical black hole.
    The pictures show trajectories of light rays 
    falling into the vortex core.
    In case the angular momentum of an incident ray
    lies within a critical interval,
    the ray is dragged by the vortex flow and finally falls into the 
    singularity. 
    (A) We have chosen several values of the angular momentum that
    cover most of the critical interval.
    (B) The chosen values of the angular momentum lie immediately above
    the negative bound of the critical interval.
    In this case a trajectory is very sensitive to small changes
    in the angular momentum.
    The radius of the limiting circle is 
    the optical Schwarzschild radius.
    }
    \label{fig:figure2}
  \end{center}
\end{figure}

\subsection{Scaling}

A vortex flow can cause an optical Aharonov--Bohm effect and may
force light to fall into an optical black hole.
Note that these phenomena depend, in principle, entirely on the mere
presence of the vortex and not on the particular value of the
vorticity, even when ${\cal W}$ is very small.
To see this we study the scaling properties of the relativistic vortex. 
A closer inspection to Eqs.\ (\ref{ur}-\ref{gammavortex})
reveals that the dimensionless velocity profile ${\bf u}/c$
depends only on the combination $(c/{\cal W})\,r$.
Let us introduce the new variables
\begin{equation}
\bar{t} = \frac{c^2}{\cal W}\, t \,\,,\,\,
\bar{r} = \frac{c}{\cal W}\, r   \,\,,\,\,
\bar{l} = \frac{c}{\cal W}\, l   \,\,,\,\,
\bar{S} = \frac{c}{\cal W}\, S   \,\,.
\label{scaling}
\end{equation}
Due to the scaling of ${\bf u}/c$ the scaled eikonal $\bar{S}$
is a solution of the Hamilton--Jacobi equation (\ref{hjexplicit})
in the variables (\ref{scaling}).
The scale of $\bar{l}$ and $\bar{S}$ was chosen such that 
$\bar{S}$ corresponds to the explicit solution (\ref{rels})
with the parameters $\omega$ and $l$.
Note that the Hamilton--Jacobi equation (\ref{hjexplicit})
in the dimensionless coordinates $\bar{t}$, $\bar{r}$, 
and $\varphi$ does not contain ${\cal W}$ anymore.
Consequently, 
the relativistic effects on light at a vortex flow do not
disappear when the vorticity approaches zero.
However, the smaller the vorticity is
the higher should be the frequency of light, in order
to produce an equivalent Aharonov--Bohm phase shift (\ref{abphase}).
This is caused by the scaling of the eikonal $S$ that governs
the phase profile of the optical field.
Furthermore, for a low vorticity 
the optical Schwarzschild radius is accordingly small.
In our idealized model, any vortex is an optical black hole,
irrespective of the value of ${\cal W}$.
In practice, the incident light is more likely to
hit an obstacle that we have entirely ignored --- the vortex core.
Similar to a star that acts as a black hole when the
gravitational Schwarzschild radius exceeds the star's size,
a vortex is an optical black hole only if the core is 
smaller than the optical Schwarzschild radius.


\section{Summary}

A moving dielectric appears to light as an effective gravitational
field \cite{Gordon}.
At low flow velocities the dielectric acts on light in the same way
as a magnetic field acts on a charged matter wave \cite{Hannay,CFM}.
The flow plays the role of the vector potential.
We have shown how the two effective models are related to each other:
The covariant wave vector corresponds to the canonical momentum
of a light ray,
whereas the contravariant wave vector plays the role of the
kinetic momentum. 
In curved space--time, and hence in a dielectric,
co- and contravariant vectors are distinct.
For low medium velocities they differ in precisely the same way
as the canonical momentum of a charged particle 
differs from the kinetic one by the vector potential.
Additionally, we have derived and studied an effective Hamiltonian for
ray propagation in moving media.
Within the limits of geometrical optics,
ray trajectories do not depend on the polarization.
Introducing a fictitious rest mass for light we have constructed
a corresponding Lagrangian.
This way has lead us to the description
of a medium in terms of a metric \cite{Gordon}.

First--order optical phenomena are certainly detectable using 
modern interferometry and ordinary dielectric fluids.
In this way one could infer the velocity profile of
transparent incompressible liquids from interferometric measurements.
To observe some spectacular relativistic effects,
one could employ highly dispersive quantum dielectrics \cite{Lene}.
One can demonstrate an optical Aharonov--Bohm effect
and create an optical black hole 
with a quantum vortex as the center of attraction \cite{LPBH}.
Our paper has established a consistent yet simplified model
of such phenomena.
We have thus reason to hope that we may stimulate experiments
to demonstrate gravitational effects on Earth that
usually belong to the realm of astronomy.

\section*{Acknowledgements}

We got interested in the optical Aharonov--Bohm effect when U.L. visited
the University of Bristol in fall 1997.
We thank 
Daniel Andre,
Sir Michael Berry,
Balasz Gyorffy,
John Hannay,
Jon Keating,
Susanne Klein,
and
Duncan O'Dell
for their hospitality and for fruitful and pleasant conversations.
We are grateful to Harry Paul for a helpful correspondence 
and Benita Finck von Finckenstein,
Michael Nieto,
and Martin Wilkens
for conversations on the Aharonov--Bohm effect.
We thank 
Salvatore Antoci,
Carsten Henkel,
Bj\"{o}rn Hessmo,
Daniel James,
Gerd Leuchs,
Rodney Loudon,
Peter Milonni,
Wolfgang Schleich,
and 
Stig Stenholm
for discussions on the optics of moving dielectrics
and on related subjects.
U.L. gratefully acknowledges the support of 
the Alexander von Humboldt Foundation
and of the G\"oran Gustafsson Stiftelse.
P.P. was partially supported by the research consortium 
{\it Quantum Gases}
of the Deutsche Forschungsgemeinschaft.

\end{document}